\documentclass[12pt]{article}

\usepackage{booktabs} 
\usepackage{mathtools}
\usepackage{paralist}
\usepackage{graphicx}
\usepackage{amsfonts}
\usepackage[utf8]{inputenc}

\newcommand{\citep}{\cite}
\begin{document}

\title{A Method for Measuring Network Effects of One-to-One Communication Features in Online A/B Tests}

 \author{
 	Guillaume Saint-Jacques\\
 	\texttt{gsaintjacques@linkedin.com}
 	\and
 	James Eric Sorenson\\
 	\texttt{jasorenson@linkedin.com}
 	\and
 	Nanyu Chen\\
 	\texttt{nchen@linkedin.com}
 	\and
 	Ya Xu\\
 	\texttt{yaxu@linkedin.com}
 }

 \maketitle

\begin{abstract}
A/B testing is an important decision making tool in product development because can provide an accurate estimate of the average treatment effect of a new features, which allows developers to understand how the business impact of new changes to products or algorithms. However, an important assumption of A/B testing, Stable Unit Treatment Value Assumption (SUTVA), is not always a valid assumption to make, especially for products that \textit{facilitate} interactions between individuals. In contexts like one-to-one messaging we should expect network interference; if an experimental manipulation is effective, behavior of the treatment group is likely to influence members in the control group by sending them messages, violating this assumption. In this paper, we propose a novel method that can be used to account for network effects when A/B testing changes to one-to-one interactions. Our method is an edge-based analysis that can be applied to standard Bernoulli randomized experiments to retrieve an average treatment effect that is not influenced by network interference. We develop a theoretical model, and methods for computing point estimates and variances of effects of interest via network-consistent permutation testing. We then apply our technique to real data from experiments conducted on the messaging product at LinkedIn. We find empirical support for our model, and evidence that the standard method of analysis for A/B tests underestimates the impact of new features in one-to-one messaging contexts.

\end{abstract}

\section{Introduction}
Many social networks offer messaging features, i.e. a way for users of the platform to communicate with each other. In the routine development of such features, A/B testing is often used as a way to gauge the effect of changes: does changing the interface lead to users sending each other more messages? Does offering AI-driven smart replies lead users to interact more? Does the availability of emoji change the way users communicate? 

A/B testing is a simple comparison of two different experiences (A and B). In its typical implementation, A/B testing relies on Bernoulli randomization: each user of the platform has an equal probability $p$ of being selected to be part of the experiment, and all selected users receive the new feature, to form the treatment group. Users who do not receive the features are called the control group. Most platforms then analyze the performance of these two groups on some outcomes of interest using a two-sample t-test. If the outcome of interest was the number of messages sent by users, the t-test would compare the average number of messages sent by treated members to the average number of messages sent by control  members. The result of this comparison gives a direct point which can be used for decision making.

However, this method relies on the assumption that the control member's outcomes are the same that they would be if the feature did not exist at all. That is, a control member's behavior is in no way polluted by the fact that other members are treated with a new feature. This assumption is known as the Unit Treatment Value Assumption (SUTVA) \cite{Rubin1986Statistics}. If that assumption is violated, there is said to be interference \cite{rosenbaum_interference_2007}, or a network effect, and the A/B test no longer yields correct results. Specifically, when SUTVA is violated, A/B test results no longer provide an accurate measurement of the impact of a new feature on the ecosystem.

A stylized example can make this clear: imagine a feature that doubles the number of messages sent by \textit{all} users, as long as at least one user has that feature available. First, consider the real effect of this imaginary feature on the total number of messages sent on our platform. Prior to the experiment, users send $M$ messages per user before the feature is introduced. Once the feature is introduced, this becomes $2M$ per user. Now, consider what a standard A/B test would measure: On the treatment side, all users send $2M$ messages. On the control side, because of the influence of treatment, all users also send $2M$ messages. The A/B testing platform computes the difference, and estimates that the effect of treatment is zero, which is very far from the true effect we described above.  

In the context of messaging features, it is clear that we should worry about network effects. For example, if a new feature leads users to send more messages, many of these messages will be sent to recipients who are themselves in the control group. These recipients will respond to these messages, sending more messages than they would if no one had the feature available. There are a number of methods known for correcting for network effects in experiments. These methods include techniques for altering the experimental design \cite{Ugander_kdd_cluster_13} or analysis \cite{aronow_estimating_2017}, but the there is not a single clear method for accounting for network effects - the appropriateness of any given technique for depends on the properties of the network and the hypotheses being tested \cite{eckles_design_2017}.

At LinkedIn, our most common cluster-based approaches \cite{egoclusters} require prior knowledge of the network structure ahead of time. However, LinkedIn is frequently used by our members to reach out to find new opportunities and to communicate and connect to grow their network. It is rarely the case that structure of a network composed of one-to-one messages is easily predicted ahead of time. In other words, it is not possible to know which pair of users will send messages to each other on the platform before the start of the experiment. Thus cluster based randomization was not an option. Moreover, other methods of correcting for network effects via adhoc analysis are concerned with correcting for a hierarchical network structure \cite{aronow_estimating_2017} that is again, quite unlike what we find in a one-to-one messaging context.

In this paper, we illustrate a simple method to account for some of these effects and obtain a corrected A/B test result without needing to change the Bernoulli randomization in the design stage. The method we propose applies to messaging, but also to any one-to-one communication feature a platform might have. We do so by leveraging a distinguishing feature of messaging networks: they are often one-to-one, meaning that we can count messages that are crossing the treatment/control boundary.

We first review the relevant literature on AB testing, and particularly how one-to-one communication violates core assumptions of A/B testing. We then propose a novel method for resolving network effects in one-to-one messaging scenarios. We review network effects in a messaging context, and how use of edge based contrasts might be used for treatment estimation. We develop a model for estimating effects of interest, especially the average treatment effect, and review our method for variance estimation - network consistent permutation testing. We next introduce several approximations for our effects of interest. Finally, we review empirical results from applying our method to real-world experimentation data from LinkedIn.

\section{Literature Review}

In this section we review the theoretical foundations and
applications of A/B testing as well as design and analysis methods for measuring network effect.
The theory of controlled experiment dates back to Sir Ronald A.
Fisher's experiments at Rothamsted in the 1920s \citep{yates1964sir}. Since then, a large variety of literature from different fields have enhanced the theoretical
foundations \citep{box2005statistics,gerber2012field} for running controlled experiments, including the widely adopted Rubin-Neyman Causal Model \citep{rubin1974estimating}. A/B testing has also become widely adopted as the golden standard for evaluating new product ideas across many industries. Previous work \citep{tang2010overlapping, kohavi2013online, bakshy2014designing, xu2015} has described some of the in-house solutions built at large companies for designing, deploying and analyzing complex experiments at scale. The best practices and pitfalls for online experiments \citep{kohavi2012trustworthy, dmitriev2017dirty, Kohavi2009, xu2018sqr}, as well as specific related topics such as variance reduction \citep{deng2013improving}, heterogeneous treatment effect detection \citep{wager2017estimation,athey2016recursive, xie2018false} and invalid experiments diagnosis  \citep{invalidity} have also been described. Our work adds to this literature by outlining a technique for accounting for network effects in A/B testing.

To set the notation and context, we review the Rubin-Neyman Causal Model \cite{rubin1974estimating}, a widely used framework to estimate causal effect. Let $M_i$ be the outcome variable for user i, e.g. number of messages sent. Let $Z_i \in {0, 1}$ be the treatment assignment of user i, where $Z_i = 1$ if that user is in treatment, and $Z_i = 0$ otherwise. Under the Rubin Causal framework, each user has two potential outcomes:
\begin{equation}
    M_i = m_i^1 \indent if \indent Z_i = 1 \indent \forall i = 1,...,n
\end{equation}
\begin{equation}
    M_i = m_i^0 \indent if \indent Z_i = 0 \indent \forall i = 1,...,n
\end{equation}
We are interested in estimating the Average Treatment Effect
(ATE), the difference of the average outcomes between applying the
treatment experience to the \textit{entire} user population and relative to applying control experience to the \textit{entire} user population. By definition, the difference or $\Delta$ between the observed outcomes of users treatment and control is
\begin{equation}
    \Delta = \sum_{i=1}^nm^1_i - \sum_{i=1}^nm^0_i
\end{equation}
In practice, an average treatment effect expressed as a percentage change from baseline is preferred for interpreting business impact.
\begin{equation}\label{eq:typical lift}
    \Delta\% = \frac{\sum_{i=1}^nm^1_i - \sum_{i=1}^nm^0_i}{\sum_{i=1}^nm^0_i}
\end{equation}
While $\Delta\%$ is unobservable, it is often estimated by

\begin{equation}
    \hat{\Delta\%} = 
    \frac{\frac{1}{n_T}\sum_{i, Z_i=1}M_i - \frac{1}{n_C}\sum_{i, Z_i=0}M_i}{\frac{1}{n_C}\sum_{i, Z_i=0}M_i}
\end{equation}
where $M_i$ is the observed value for individual i under a randomized experiment and $n_T$, $n_C$ represent the number of units in treatment and control. $\hat{\Delta\%}$ is asymptotically unbiased for estimating 
$\Delta\%$ provided the Stable Unit Treatment Value
Assumption (SUTVA) holds. SUTVA states that the behavior of each user in the experiment depends only on his own treatment and not on the treatments of others.

However, such assumption may not hold when testing new features in a social context such as messaging. Violation of SUTVA, which is commonly known as interference \cite{rosenbaum_interference_2007}, leads to biased estimation of treatment effect. The most common approach to improve the estimate when SUTVA is violated is Cluster-based randomization \citep{aronow2013class,eckles2017design,ugander2013graph,pouget2018optimizing}, where
users are clustered based on the user-to-user interaction graph (e.g connections) and treatment is
assigned at a cluster level.
Alternative methods include Multi-level designs \citep{hudgens2008toward,tchetgen2012causal}, where randomization is still on individual level but treatment is assigned to different proportions, as well as model based method \cite{basse2015model,gui2015network}, where, as an example, the interference effect is assumed to be proportional to the number of neighbors treated.
Many of these methods require the network to be observable or predictable and do not work well in a simple Bernoulli randomized experiment.

\section{Network effects in one-to-one messaging applications }\label{sec:networkEffects}
 

A specific feature of messaging experiments is that the main behavior of interest that is impacted by the experiment - the sending of messages - typically only involves two individuals: the message sender, and the recipient. For each message sent, we can retrieve the treatment status of the sender (i.e. treated or control), as well of the treatment status of the recipient. 


\begin{figure}[h!]
	\includegraphics[width=35mm]{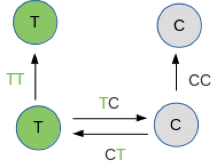}
    \caption{A taxonomy of message classes by treatment status of sender and recipient}
    \label{fig:tcIllustration}
\end{figure}

This feature of the problem naturally allows us to classify messages into four categories (Figure \ref{fig:tcIllustration}): messages from a treated member to another one (denoted of class TT), messages from a treated member to a control member (class TC), from a control member to a treated member (class CT), and messages within the control group (class CC). Throughout this paper, we will show that with a few assumptions and proper normalization, functional forms of the total value of these classes can yield insights into potential network effects. To establish the notation for later sections, in an experiment with members $i = 1, ..., n$, we have
\begin{equation} \label{eq:contrasts}
    M_{TT}:=\sum_{i\in T, j\in T }M_{ij}, \indent 
    N_{TT} = \sum_{i, j}\mathbb{I}_{i\in T, j\in T}
\end{equation}
And similarly, we can define $M_{TC}, M_{CT}, M_{CC}$ and $N_{TC}, N_{CT}, N_{CC}$.

Such a classification can be performed for any type of experiment. For the purposes of this paper, we focus on Bernoulli randomizations, which are the most common and typically the easier to set up using most experimentation platforms. Our main purpose is to use the above classification, paired with the simplest possible randomization technique, to estimate the network effect, rather than rely on non-standard randomization techniques, such as edge-level randomization or cluster-based randomization. As a simple example, if treatment is Bernoulli-randomized to 50\% of members, and the treatment is a placebo, one would expect, over all possible Bernoulli randomization of users, the number of edges and the number of messages to be equal across all classes:
\newcommand{\E}{\mathbb{E}}
\begin{equation}
    \E(M_{TT})=\E(M_{TC})=\E(M_{CT})=\E(M_{CC}) \footnote{The expectation is taken over all possible Bernoulli randomization of users}
\end{equation}
\begin{equation}
    \E(N_{TT})=\E(N_{TC})=\E(N_{CT})=\E(N_{CC}) = \frac{n^2}{4}
\end{equation}

However, if the experiment has uneven traffic between treatment and control, the class totals will not be equal since the number of edges in each class will be different and one needs to apply normalization accordingly.

Note that in Bernoulli randomized experiments with treatment probability $p$, and the probability of control is $1-p$.
\begin{equation}
    \mathbb{E}(\frac{N_{TT}}{n^2}) = p^2, \mathbb{E}(\frac{N_{TC}}{n^2}) = \mathbb{E}(\frac{N_{CT}}{n^2}) = p(1-p),
    \mathbb{E}(\frac{N_{CC}}{n^2}) = (1-p)^2
\end{equation}

\paragraph{Placebo treatment} \label{placebo}
In this case, we expect that, for any ramp percentage (i.e. for any proportion of users treated), the per-edge number of messages is identical: 
\begin{equation}
    \E(\frac{M_{TT}}{N_{TT}}) = \E(\frac{M_{TC}}{N_{TC}})
    = \E(\frac{M_{CT}}{N_{CT}}) = \E(\frac{M_{CC}}{N_{CC}})
\end{equation}
\begin{equation}
    \frac{(1-p)^2}{p^2}\E(M_{TT}) = \frac{1-p}{p}\E(M_{TC})
    = \frac{1-p}{p}\E(M_{CT}) = \E(M_{CC})
\end{equation}

\paragraph{No response}\label{noresponse}
In the case of messaging experiments, it is in general interesting to find out whether the new messages users are sending due to an experiment also generate responses. If the primary impact of treatment is to have users send more messages, then one would expect the CT class to contain some messages that were created as a response to messages received from a treated member (TC), but the CC class should not contain any. Therefore, one can look at the following response contrast: 
\begin{equation}
    \frac{1-p}{p}\E(M_{CT}) = \E(M_{CC})
\end{equation}
 
If that contrast is non-zero, then it is likely that treatment has an impact on the number of messages sent by control members. Note that pollution of the control group (i.e. the alteration of control group users due to the existence of treatment) is a primary cause of network effects, and therefore should be addressed.

\paragraph{No treatment affinity}\label{noaffinity}
A potential feature of an intervention is that treated users may preferentially target other treated members. This causes problem with regular A/B testing inference too, as it means that control members receive relatively fewer messages than they would if the treatment did not exist. A useful contrast to capture this is the affinity contrast: 

if $p=\frac{1}{2}$, affinity contrast $= \E(M_{TT}) - \frac{1}{2}\E(M_{TC}+M_{CT})$
 
in general, affinity contrast $ =\frac{(1-p)^2}{p^2} \E(M_{TT}) - \frac{1-p}{2p}\E(M_{TC}+M_{CT})$

The intuition behind this contrast is as follows: treated members will send more messages to other treated members, and these messages generate response, which is all captured under $M_{TT}$
. They also send more messages to control members. These messages are captured in $M_{TC}$, and the responses they generate are captured in $M_{CT}$. Assuming treated and control members have the same response rate to these new messages, then a properly normalized difference of these terms should be equal to zero. 

\paragraph{Perfect treatment affinity}\label{perfectaffinity}
Some specific experiment may also induce treated members to send more messages to other treated members only. In this case, one would expect to see (in the case of a positive treatment effect): 
\begin{equation} \label{eq:perfectAffinity}
    \frac{(1-p)^2}{p^2} \E(M_{TT}) > \frac{1-p}{p}\E(M_{TC}) = \frac{1-p}{p}\E(M_{CT})=\E(M{CC})
\end{equation}

\subsection{Potential outcomes used for identification }

We now formalize the intuition behind the above contrasts using potential outcomes. 

For each variable $X$:
\begin{itemize}
\item $X^1$ is the value of the contrast that is observed (i.e. with the treatment variant being different form the control variant). 
\item $X^0$ is the counterfactual value of the contrast if treatment is a dummy (the treatment variant is identical to the contrast variant. 
\end{itemize}

So for example $M^1_{TT}$ denotes the observed value of 
$M_{TT}$ whereas $M^0_{TT}$ denotes the value $M_{TT}$ would take if treatment had not been rolled out. 

With this notation in mind, we can set up some counterfactuals, and  reformulate the above contrasts: 
if $p = \frac{1}{2}$, $M^0_{TT} = M^0_{TC} = M^0_{CT} = M^0_{CC} $.
For all other p: 
\begin{equation} \label{e1}
    \frac{(1-p)^2}{p^2} M^0_{TT} = 
    \frac{1-p}{p}M^0_{TC} = 
    \frac{1-p}{p}M^0_{CT} = 
    M^0_{CC}(1)
\end{equation}
A crucial assumption we add is that the control-control edges are undisturbed by the existence of treatment: 
\begin{equation}
    M^1_{CC}(p) = M^0_{CC}(p) = M^0_{CC}(0)\indent \forall{p}\in [0,1]
\end{equation}
We further assume that our properly normalized M totals are independent of the ramp percentage: 
\begin{equation}
    M^1_{TT}(p) =  M^1_{TT}(1)\indent \forall{p}\in [0,1]
\end{equation}

We will discuss how well these assumptions hold in one-to-one communications based on real experimental data in section 7.
The main outcome of interest for any messaging experiment is the total of messages that would be sent if treatment were rolled to 100\% of users: $M^1_{TT}(1) - M^0_{TT}(0)$

\section{The Model}
\subsection{Estimating the Total  Treatment Effect}\label{sq:ate}
 In order to measure the total effect of a new messaging features, we need to estimate the number of messages that would be sent if all users were treated, as opposed to if none were $M^1_{TT}(1) - M^0_{TT}(0)$.
 Using equation \ref{e1} and substituting, we can express the total effect as a function of observables only:  
 \begin{equation}
  M_{TT}^1(1)-M_{TT}^0(1)= \frac{1}{p^2}M_{TT}^1(p)-\frac{1}{(1-p)^2}M_{CC}^1 (p) 
 \end{equation}
 
\subsection{Estimating the experiment-specific response rate}
Different messaging experiments may induce members to send different types of additional messages; some may be more likely to elicit responses from their recipients then others. While not indispensable to identifying the total treatment effect, it can be helpful to try and estimate the experiment-specific response rate. The intuition behind our approach is as follows: treated members will send more messages to the control group, which will affect the total number messages of the TC class. In return, control members will respond to these messages, which will affect the number of messages of the CT class. Despite this happening. None of this should affect the number of messages in the CC class. In other words, one can interpret a normalized difference of messages between TC and CC as the number of "extra messages" sent from the treatment group to the control group, and a normalized difference of messages between CT and CC as the number of "extra responses" that were created as a result of these incoming messages. Therefore, we want to estimate the ratio of these extra responses over these extra incoming messages to the treatment group: 
\begin{equation}
    \alpha = \frac{M^1_{CT} - M^0_{CT}}{M^1_{TC} - M^0_{TC}}
\end{equation}
Reformulating this as a function of observables only: 
\begin{equation} \label{eq:elementAlpha}
    \alpha = \frac{M^1_{CT} - \frac{p}{1-p}M^1_{CC}}{M^1_{TC} - \frac{p}{1-p}M^1_{CC}}
\end{equation}
 
\subsection{Estimating "instant lift"}
While also not necessary to correctly estimate the total treatment effect, in some cases, it may also be useful to tell apart messages that are created as a direct consequence of treatment on treated members from messages that appear as responses to these messages. This requires some more modelling assumptions.  

In the simplest formulation, one can decompose the problem as follows:

\begin{itemize}
\item Call $q_1$ the instant lift on treated members. This is the instant lift of messages from treated members to other treated members. Similarly, call $q_2$ the instant lift of messages of treated members to control members. Differences between $q_1$ and $q_2$ are referred to as \emph{affinity}, i.e. treated members showing different behavior towards control versus treated members. Upon being treated, member's number of sent messages is multiplied by $1+q_1$ 

\item Assume that messages have a constant (though specific to each experiment) probability of generating a response, called $\alpha$. 

\end{itemize}

Then, the total treatment effect is simply 
\begin{equation}
 q_1(1 + \alpha + \alpha ^2 ...) = \frac{q_1}{1-\alpha}
\end{equation}
This can straightforwardly be estimated as a function of observables:
\begin{equation} \label{eq:elementLift}
    M_{TT}^1 = \frac{q_1}{1-\alpha}M_{TT}^0 = \frac{q1}{1-\alpha}
    \frac{p^2}{1-p^2}M_{CC}^1
\end{equation}
Substituting our estimated $\alpha$ as well as observed totals for $M^1_{TT}$ and $M^1_{CC}$ then yields an estimate for the total lift. 

\section{Variance estimation with network-consistent permutation tests on spark } \label{sec:varianceEst}

An important part of any experiment is gauging the significance of the results. Our experiment result data has the following structure: 

\begin{itemize}
    \item \textit{src (int)}, memberId of sender
    \item \textit{dest (int)}, memberId of recipient
    \item \textit{msg (int)}, number of messages
    \item \textit{srcT (boolean)}, treatment status of sender (1=treated)
    \item \textit{destT (boolean)}, treatment status of recipient
\end{itemize}

We use the permutation method for generating non-parameteric confidence intervals. Since we are using edge-level totals, we leverage three different types of permutations: 

\begin{itemize}
    \item \textbf{Full permutation}, where for each edge, both sender and recipient treatment status is shuffled 
    \item \textbf{Sender-side permutation}, where the treatment status of recipient is retained, and status of the sender is shuffled 
    \item \textbf{Recipient-side permutation}, which does the opposite
\end{itemize}

Crucially, these three edge-level permutations are done in a network-consistent manner, meaning that within a permutation iteration, the treatment assignment of a specific user is kept consistent. For example, if user 123 is classified as a treated sender in iteration 1, she is classified as such in all edges where she is a sender in iteration 1. Her assignment to treatment in iteration 1 does not influence her assignment in subsequent iterations.

To do this practically in Spark, we make heavy use of hash functions, as it is not feasible to store treatment status of each member for each iteration. Rather, it has to be recomputed on the fly. A random seed is concatenated with the member's id and the id of the current permutation iteration (for example, from 1 to 10,000), to give a random treatment/control assignment that is consistent on all edges the member is involved in, without communication between spark executors (in practice, we use MurMurHash3 to perform the hashing). A brief chart with the most relevant Scala code is given in Figure \ref{fig:algo} of the appendix.

The different permutation methods allow us to test different null hypotheses. For example, if we want to test whether the intervention has any effect at all (i.e. whether it's a placebo treatment 
, we can use the full permutation. The logic is that if there is no treatment effect whatsoever, \textit{the treatment assignment label (treatment/control) of senders and recipients should not matter}. However, if we want to allow for the fact that the treatment can cause treated members to send more messages, but test for other effects, full permutation is no longer appropriate. For example, if we want to test whether control and treated members \textit{receive} a different number of messages, then we should use destination-side randomization: \textit{the label of senders is expected to matter, but the label of recipients should not matter}. 

\section{Approximations of effects based on common summary statistics}
In the previous sections, we discussed how to estimate the total treatment effect that captures both the "instant lift" and the lift due to response based on edge-level contrasts. In this section, we offer an approximation mechanism which is based on summary statistics that are usually readily available in an A/B testing system. 

Without loss of generality, a Bernoulli randomized experiment has one treatment (T) with traffic proportion p and one control (C) with traffic proportion (1-p). For an edge-level action with an actor and receiver, both the difference between the two groups on the actor side and the receiver side can be measured. Using messages sent between members of the network as an example, these lifts would be computed as: 

\begin{equation}
    \Delta\%_{Send} = \frac{MS_T/n_T}{MS_C/n_C} - 1
\end{equation}
\begin{equation}
    \Delta\%_{Receive} = \frac{MR_T/n_T}{MR_C/n_C} - 1
\end{equation}
where 
\begin{itemize}
    \item $MS_{T} = \sum_{i \in T}\sum_{j = 1}^n M_{ij}$, 
    $MS_{C} = \sum_{i \in C}\sum_{j = 1}^n M_{ij}$ are total number of messages sent by treatment and control groups, respectively. 
    \item $MR_{T} = \sum_{j \in T}\sum_{i = 1}^n M_{ij}$,
    $MR_{T} = \sum_{j \in T}\sum_{i = 1}^n M_{ij}$
    are total number of messages received by treatment and control groups, respectively
    \item $n_T, n_C$ are sample sizes of treatment and control groups, respectively.
\end{itemize}

 Note that this is the typical lift calculation used by modern A/B testing systems, and was previously referenced in equation \ref{eq:typical lift}.
 
\newtheorem{prop}{Proposition}
\begin{prop}

Let $u = \frac{q_2}{q_1} - 1 $
\begin{equation} \label{eq:liftAprox}
\Delta\%_{Send} + \Delta\%_{Receive} \approx \frac{q_1}{(1-\alpha)}
\frac{1+2pu}{1+u} 
\end{equation}
The summation of the percentage lift of sends and receives can be used to approximate the total treatment effect when the experimental effect size is not large and at least one of the following two conditions holds:
\begin{itemize}
    \item the experiment has equal traffic allocated to treatment and control
    \item there is no affinity
\end{itemize}
\end{prop}
We leave the proof of this to the Appendix.

In the above proposition, the approximation is the corrected lift $\frac{q_1}{1-\alpha}$ when the same number of users are in treatment and control \textit{i.e.}, $p=1-p $, regardless of whether there is affinity. However, this is no longer accurate when the traffic allocated to treatment and control is unequal. When treatment traffic is smaller than that of control, the approximation using the summation of the percentage lift of sends and receives will under-estimate the treatment effect when there is positive affinity, and over-estimate the treatment effect otherwise. However, such relationship is reversed when the traffic percentage in treatment is larger than 50\%. 
We have found that this is aligned with our observation that in a regular Bernoulli randomized experiment (where positive affinity is expected), the naive estimates 
$\Delta_{Sends}\%$ and $\Delta_{Receives}\%$ gets bigger as treatment ramps to higher traffic percentage. This is in fact an artifact due to not capturing the network effect and the bias depends on the ramp size.

We can also estimate the response rate of the newly created messages, as described in section 4, using similar methods. 

\begin{prop}
\begin{equation} \label{eq:alhpaAprox}
    \frac{\Delta\%_{Receive}}{\Delta\%_{Send}} \approx
    \frac{\alpha + pu\alpha}{1+pu\alpha}
\end{equation}
The ratio of the \%lift of receives and sends can be used to approximate the response rate, $\alpha$, when affinity  is reasonably small. The proof of this proposition is also captured in Appendix.
\end{prop}

\section{Empirical Application}
 We applied our methods to a set of real data from LinkedIn with several key goals in mind. Of primary interest was determining how much our estimated total treatment effects would be changed in practice by adopting these methods. We further wanted to test whether our estimates of the total treatment effect would be robust to changes in the percentage of experimental population treated in any given experiment. Additionally, we sought to see if our parameter $\alpha$ would yield insights related to the depth of exchange in one-to-one communication, and if we could find evidence in our experiments of the 'treatment affinity' we propose above.

 \subsection{Methods}
 \subsubsection{Included Data}
 We applied our methods to the results from 43 iterations of 21 distinct experiments run at LinkedIn. At LinkedIn, we frequently use a progressive roll-out of an experimental piece of code - exposing a small percentage of traffic to the new code in an initial 'iteration' and then increasing the allotment in subsequent 'iterations'. Thus a single change to product or algorithms is tested in an 'experiment' which may be in turn composed of several 'iterations'. 

Experiments included in the our dataset tested different types of product changes, including the look and feel of the messaging experience, and how messaging notifications were prioritized. Experiments were identified for inclusion by either being an important messaging experiment from 2017-2018, or were identified in an earlier internal meta-analysis as being an experiment that had a significant impact on the messages sent metric. All experiments included in the study were required to be 2 variant, i.e. have tested a single experimental change against the control or standard version of the product. Experiments were further required to have an experimental population of at least 1 million members. Moreover, iterations of experiments were only included if they had at least 3 days of data. For each experiment, a maximum of seven days of data were included in the dataset. Experiments that overlapped in time were included in these data, but importantly, overlapping experiments used an orthogonal randomization scheme.

\subsubsection{Effect Estimation}
For each included iteration of each experiment, we ran a deployed service called Element (Edge LEvel MEtric Network Test), that implements the work so far described. 
The Element service computes the edge level contrasts listed in equation \ref{eq:contrasts}, as well as other effects of interest, notably the 'Element corrected lift' described in equation \ref{eq:elementLift}, and $\alpha$, described in equation \ref{eq:elementAlpha}. Variances for each of these parameters are estimated using the network-consistent randomization described in section \ref{sec:varianceEst}. The service also computes the approximations for lift and $\alpha$ listed in equations \ref{eq:liftAprox}, and \ref{eq:alhpaAprox}, respectively. 
For the end user, ELEMENT produces a rendered report with some automatically generated commentary and insights - we have included screenshots of these reports below in the appendix \ref{fig:mainUI}. The data statistics in the results section are analyses conducted on an aggregated set of these reports. 
In this section we refer to the 'typical' or 'standard lift computation. This standard lift is the average treatment effect expressed as a percentage of change that is described above in equation \ref{eq:typical lift}. For parity, we modified Element's permutation testing capabilities to generate confidence intervals for this typical lift estimation, again using the 'full randomization' scheme.

\subsection{Results}
\subsubsection{Estimated lift is larger than standard lift}
As seen from our approximation of the estimated lift in equation \ref{eq:liftAprox}, our corrected lifts should be approximately equal to the sum of the lift to the number of messages sent and the lift to the number of messages received. This is particularly true for experiments with relatively small impacts, and for experiments where the percentage of population that is treated, $p$, is close to 50\%. In other words, we should expect our newly corrected lifts to be larger than lifts computed with the standard methodology. Empirically, this is what we observe.

In our sample of experiments, our corrected lift was consistently much larger than lift that was observed using the typical definition of lift. A pairwise t-test of corrected  (M = 1.56\%, SD =  2.20\%) and standard lifts  (M = 1.0\%, SD =  1.62\%) for included experiments at a 50\% ramp, showed that the Element corrected lift was significantly greater than the standard lift ( t(19) = 3.23, p < 0.01). Because some of the included experiments actually had a negative lift, we repeated these calculations on the absolute values of our corrected and typical lift definitions and obtained similar results.

\begin{figure}[h!]
  \includegraphics[width=0.5\textwidth]{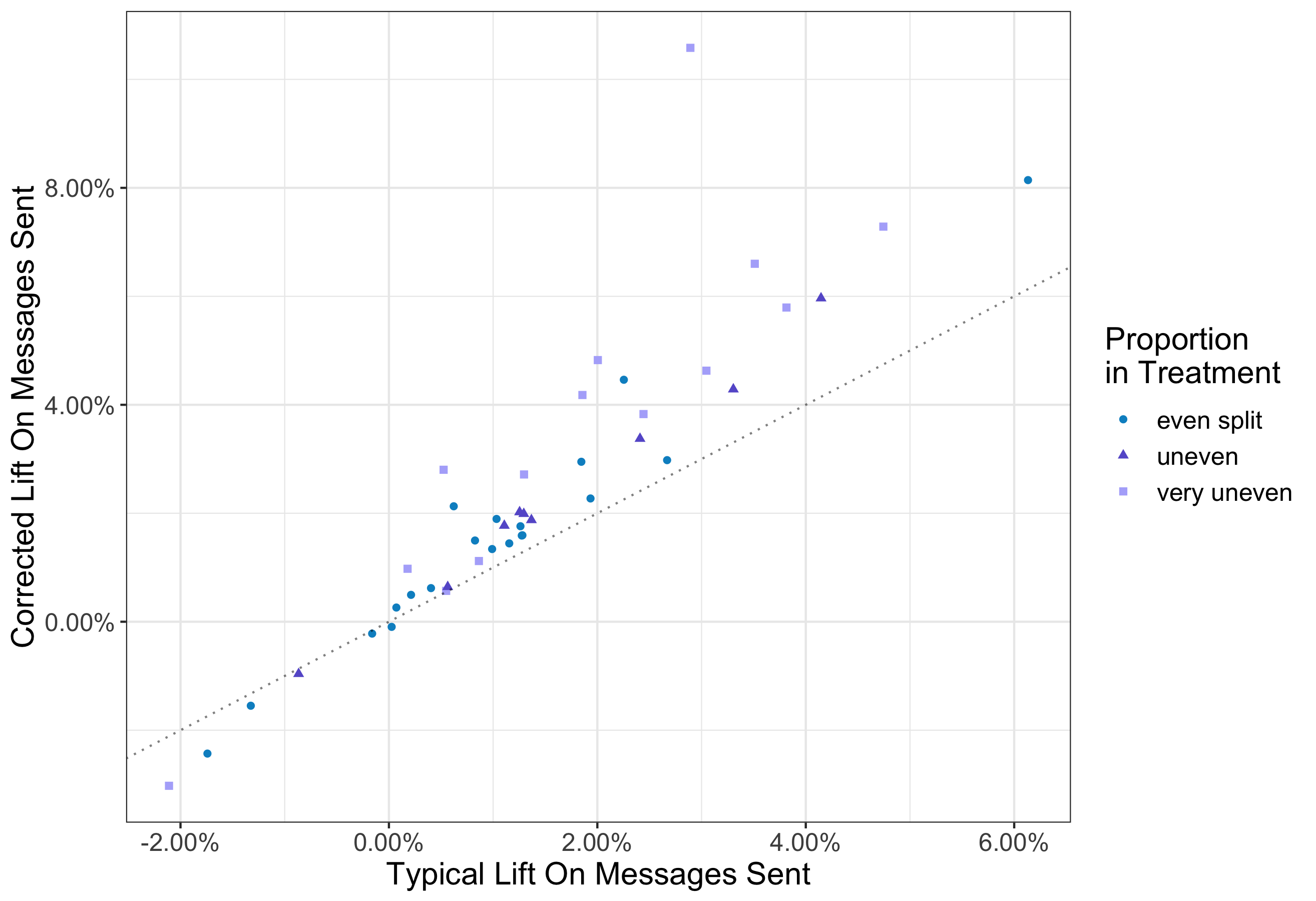}
  \caption{The plot above shows the typical computation of lift (x-axis) against the computed corrected lift (y - axis), where each point is the lift estimations for a single iteration of an experiment. The increase or decrease in estimated treatment effect can be seen by comparing the position of the points to the identity line (dotted)}
\end{figure}

Moreover, we find clear evidence that the assumptions made in computing our approximations hold quite well. To make this comparison, we binned each experimental iteration by the percentage of members in the treatment condition in that iteration. 'Very uneven' experiments had fewer than 25 \% or more than 75\% of members. 'Uneven' experiments had more than 25 \%, and less than 75\% of members in treatment, but did not have exactly 50\%. 'Even' had exactly 50\%.

\begin{figure}[h!]
  \includegraphics[width=0.5\textwidth]{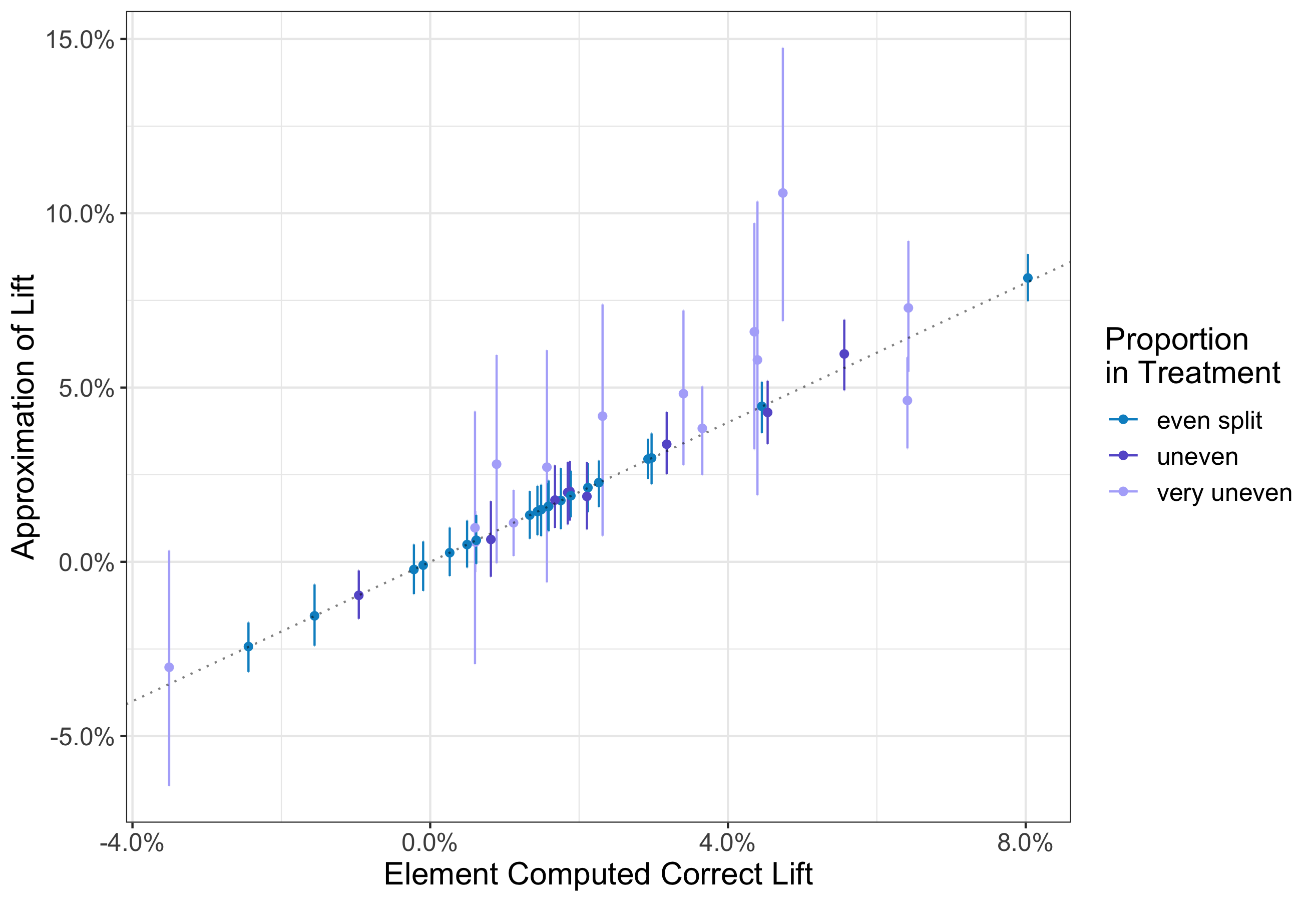}
  \caption{The plot above shows our approximations of lift (x-axis) against the computed corrected lift (y - axis). Confidence intervals for the computed lift are derived from network-consistent randomization and plotted on the y-axis. As predicted, the computed and approximated lifts match quite well for iterations with an even proportion of members in treatment and control. }
\end{figure}

\subsubsection{The total treatment effect is robust to changes in proportion of experimental population in treatment}

We previously asserted that the lift correction we are calculating compensates for the percentage of members given treatment in any iteration. If this is true, then our lift estimates should be relatively stable within the same experiment. One way to test this is to look at all experiments in our dataset that have multiple iterations, compute the point estimate and confidence interval for each iteration, and then assess if the confidence intervals of each experiment overlap with all the other confidence intervals from that experiment.

Because our data are data of convenience, this analysis does not have uniform data for all experiments. 12 individual experiments had at least 2 iterations; one of these had as many 5 different iterations. For each iteration of all 12 experiments, we calculated 90\% confidence intervals for the Element corrected lift, and the standard (send only) lift using the 'full randomization' regime described above.

 We found that 7 experiments had confidence intervals that overlapped for all iterations for both measures. 4 experiments had non-overlapping confidence intervals for both Element lift and standard lift, only one experiment had non-overlapping confidence intervals for Element lift, but overlapping confidence intervals for the standard lift. We were initially disappointed with this result, but upon close inspection, the failures of overlap highlight some important limitations of Element. 

The experiment that had overlap for standard, but not Element confidence intervals had just 1 of 4 variants that did not overlap with the others. Crucially, this iteration had only 4 days of data (the other 3 iterations had 7), and had a low percentage of members in treatment (10\%). When we considered the other 4 experiments, we found 1 experiment had an outstanding variant with just 2 days of data. The other experiments had inconsistent changes between iterations (user interface experiences or algorithmic outputs that differed between iterations). Thus, it appears that Element lift estimates are generally stable, though they may require at least 7 days of data per variant. 

\subsubsection{Alpha identifies experiments that create different 'types' of conversations}

Our measure of $\alpha$ can be interpreted as the response rate to new messages 'created' by the experiment, or a ratio of the degree to which one-to-one communications between treatment and control are dominated by treatment. We computed $\alpha$ using full randomization for all 42 individual iterations, using full randomization. 

Empirically, we see that the confidence interval for $\alpha$ includes 0 more often (24 iterations) than not (19). This lack of significance is perhaps unsurprising because our estimate is the ratio of two normalized differences - making the confidence interval generally large. Interestingly, $\alpha$ is not more likely to be significant for 50/50 ramps vs. very uneven ramps. Of the 35 ramps that had a significant effect on sends (from permutation testing), 19/35 (54\%) had a significant $\alpha$. 

Restricting our view to only those iterations with a significant $\alpha$, we find that these iterations sort into 3 different 'types' of experiment. The first type is iterations that generated 'one-off conversation' iterations. This type consisted of 9 iterations that had an $\alpha$ similar to our global 24 hour reply rate. This similarity indicated many more messages were sent from treatment to control members than from control to treatment. The second type we refer to as 'long conversation' iterations. Here, 8 iterations had an $\alpha$ much higher than our global response rate, but remained below 100\%. In these iterations, more messages were sent from treatment to control than from control to treatment, but the ratio was much closer to equal. In order to have a more equal count, conversations between members in treatment and control would have had to have been long, as the initial send from treatment did not dominate the comparison.

The final type, which was seen in two iterations, were outliers. For these iterations, response rates exceed 100\%, indicating that more complex network effects were present. Importantly, these outliers violated our assumptions in instructive ways. The first outlier was an experiment that did not alter the propensity to send messages. Instead, the experimental manipulation was a change that instead made a member more likely to receive a message. This shows that in order for the Element technique to generate interpretable outputs, the basic experimental setup must be constrained to manipulations that act primarily on the message sender. The second outlier demonstrated perfect affinity, and will be discussed at more length in the next section.

Together, these findings show that when an experiment primarily changes the initial sender experience, $\alpha$ may be used to gain some insight into the type of one-to-one interactions that a given experiment induces. This type of insight is especially useful in an A/B testing context as it can help provide decision makers more insight into why or how their new feature is changing user behavior. For example, imagine a new feature that was designed to encourage users to continue existing conversations with a lot of back-and-forth between user. If Element analysis showed this experiment had a significant increase in messages sent but low alpha, this would be an important signal to redesign the product instead of giving it to all users in the ecosystem.

\subsubsection{Perfect Affinity}
In the previous section, we mentioned the existence of a second that demonstrated perfect affinity. This iteration was one of only two iterations (both from the same product experiment), to show perfect affinity. \footnote{We also noted that $\alpha$ was not valid for one of these iterations. In fact, $\alpha$ was not valid for either iteration, but was spuriously significantly different from 0 for only one of the iterations. Inspecting our formula for $\alpha$ (equation \ref{eq:elementAlpha}), it becomes clear why we should not use $\alpha$ to make inferences in cases such as these: if there is no significant difference between the TC and CC cases, the expected value of the denominator for our estimation is 0.} The 'outlier' iteration was an iteration in which 85\% of the population were in the treatment group. The other iteration had 50\% of users in the treatment group.

The product experiment was one in which 'Active Status' was introduced to the LinkedIn messaging platform. Active Status allows members to see which of their connections are currently active and reachable on LinkedIn. It does so by showing a green dot indicator next to members connections' name if they are active and reachable.

During the A/B testing of this feature, only members in the treatment group were able to see the green dot indicators. Crucially, Active Status indicators would also only be shown for members in the treatment who had opted in to the experiment. Thus, the reason for affinity becomes clear - members in the treatment group would only have a reason to send 'extra' messages to their connections who showed a presence indicator - and these members would exclusively also be in the treatment group.

To formally test for affinity, we used edge level contrasts described in section \ref{sec:networkEffects} to compare edge types. For both affinity iterations, we used 'full randomization' to generate confidence intervals for all edge level contrasts. For both iterations, We found that the ramp percentage normalized summation of messages sent from treatment to treatment (TT), but not treatment to control (TC), or control to treatment (CT) differed significantly from the expected baseline contrast, CC. Furthermore, the TT contrast was significantly greater than the TC contrast for both iterations, indicating perfect affinity. In other words, the experiment induced members in the treatment group to send more messages than they otherwise might have, but to send these messages only to other members of the treatment group.

Perfect affinity also drove a large difference between standard lift and Element lift. This difference was also dependent upon the proportion of treated members. Using the standard lift estimation technique, we found that at the 50\% iteration, the new feature increased messages sent by +2.3\%, and increased by +3.0\% at the 85\% iteration. By contrast, the Element corrected lift was much more consistent: +4.4\% for the 50\% iteration and +4.6\% for the 85\% iteration. The standard method for estimating lift dramatically underestimated the increase in messaging caused by Active Status for both iterations. Moreover, the underestimate was largest at the 50\% iteration, the iteration that is generally supposed to have the best statistical properties for inference \cite{xu2018sqr}.

Thus, Element allowed us not only to identify the presence of perfect affinity, but to corrected for it as well. From the description of the experimental setup, it might be fairly trivial to recognize that this would be a situation that should have strong network effects. Indeed, it makes perfect sense that members in the treatment group were only encouraged to send extra messages to others in the treatment group - these were the only connections for whom they might be able to see the Active Status indicator. By using Element we were able clearly confirm these intuitions. Perhaps more importantly, Element was able to produce a corrected lift estimate at the 50\% iteration that held true at higher ramp percentages. 

This example highlights our method's major benefits. The method can identify the presence of strong network effects like perfect affinity, and produce a corrected estimate with just a single standard 50\% Bernoulli randomized iteration, obviating the need for more costly experimental designs or iterations. Further, it improves both the quality and speed of decision making when A/B testing new products or features that impact the one-to-one interactions of users.

\bibliographystyle{acm}
\bibliography{element} 

\newpage
\section*{Appendix}
\subsection*{Approximation for treatment effect}\label{appen:approximation}
Here we show that 
\begin{equation*} \label{eq:liftAprox2}
\Delta\%_{Send} + \Delta\%_{Receives} \approx \frac{q_1}{(1-\alpha)}
\frac{1+2pu}{1+u} 
\end{equation*} and 
\begin{equation*} \label{eq:alhpaAprox2}
    \frac{\Delta\%_{Receives}}{\Delta\%_{Send}} \approx
    \frac{\alpha + pu\alpha}{1+pu\alpha}
\end{equation*}
As discussed in Section 4, we can denote the contrasts in terms of $q_1$, $q_2$, $\alpha$ as following

\begin{equation*}
    \frac{M_{TT}^1}{p^2} = \frac{q_1}{1-\alpha}C
\end{equation*}
\begin{equation*}
    \frac{M_{TC}^1}{p(1-p)} = \frac{q_2}{1-\alpha^2}C
\end{equation*}
\begin{equation*}
    \frac{M_{CT}^1}{p(1-p)} = \frac{q_2\alpha}{1-\alpha^2}C
\end{equation*}

where $C = \frac{M_{CC}^1}{(1-p)^2}$.

Note also that $MS_T$, $MS_C$, $MR_T$, $MR_C$, by expectation, are weighted linear sums of the contrasts above.
\begin{equation*}
\frac{MS_T}{p} = p\frac{M_{TT}^1}{p^2} + (1-p)\frac{M_{TC}^1}{p(1-p)} =\frac{p(1+u)q_2(1+\alpha) + (1-p)q_2}{1-\alpha^2}C
\end{equation*}
\begin{equation*}
\frac{MS_C}{1-p} = p\frac{M_{CT}^1}{p(1-p)} + (1-p)\frac{M_{CC}^1}{(1-p)^2} = (\frac{pq_2\alpha}{1-\alpha^2}+1)C
\end{equation*}
\begin{equation*}
\frac{MR_T}{p} = p\frac{M_{TT}^1}{p^2} + (1-p)\frac{M_{TC}^1}{p(1-p)} =\frac{p(1+u)q_2(1+\alpha) + pq_2\alpha}{1-\alpha^2}C
\end{equation*}
\begin{equation*}
\frac{MR_C}{1-p} = p\frac{M_{TC}^1}{p(1-p)} + (1-p)\frac{M_{CC}^1}{(1-p)^2} =(\frac{pq_2}{1-\alpha^2}+1)C
\end{equation*}

Hence
\begin{equation*}
\Delta_{Sends}\% = \frac{\frac{MS_T}{p}}{\frac{MS_C}{1-p}} - 1 = \frac{q_2(1+up(1+\alpha))}{1-\alpha^2 + pq_2\alpha} =
\frac{q_2(1+up(1+\alpha))}{1-\alpha^2} + o(\frac{1}{q_2^2})
\end{equation*}
The last equation can be derived using Taylor expansion around $q_2 = 0$.

Similarly,
\begin{equation*}
\Delta_{Receives}\% = \frac{\frac{MR_T}{p}}{\frac{MR_C}{1-p}} - 1 = \frac{q_2(\alpha+pu(1+\alpha))}{1-\alpha^2 + pq_2\alpha} =
\frac{q_2(\alpha+pu(1+\alpha))}{1-\alpha^2}  + o(\frac{1}{q_2^2})
\end{equation*}

Finally, when $q_2$ is not too big,
\begin{equation*}
\Delta_{Sends}\% + \Delta_{Receives}\% \approx \frac{q_2(1+2pu)}{1-\alpha} = \frac{q_1}{(1-\alpha)}
\frac{1+2pu}{1+u} 
\end{equation*}

and 
\begin{equation*} 
    \frac{\Delta\%_{Receives}}{\Delta\%_{Send}} \approx
    \frac{\alpha + pu\alpha}{1+pu\alpha} \approx \alpha 
\end{equation*}
Again, the last equation can be viewed through Taylor expansion around u = 0

\newpage

\begin{figure*}[h!]
	\includegraphics[width=0.45\textwidth]{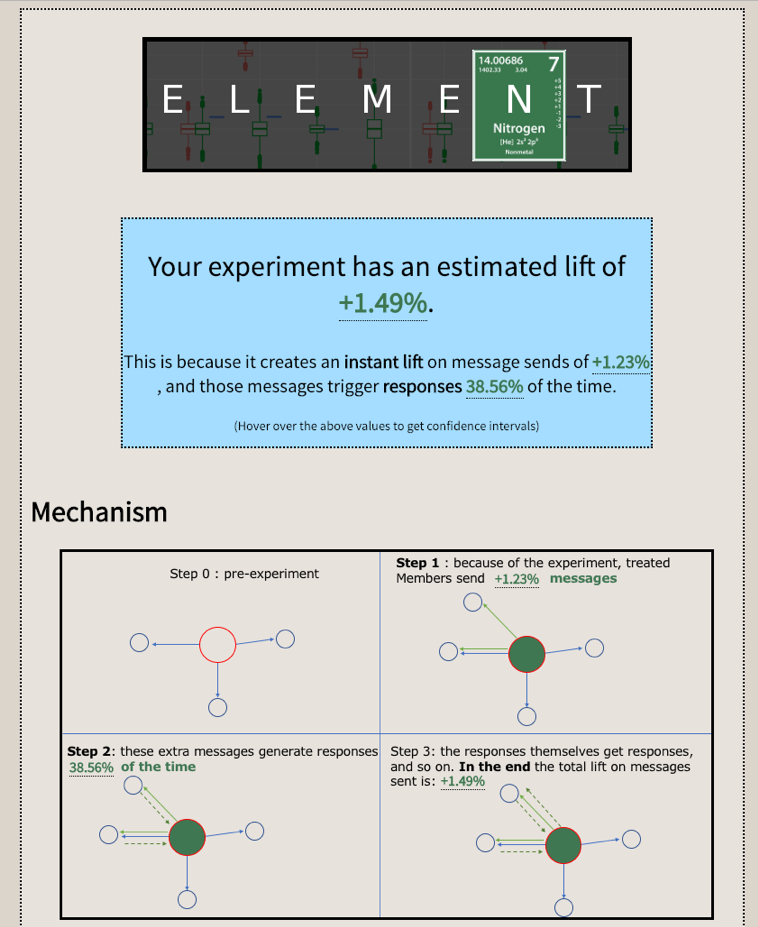}
    \caption{Main user interface displaying the results in a user-friendly way}
    \label{fig:mainUI}
\end{figure*}

\begin{figure*}[h!]
	\includegraphics[width=0.45\textwidth]{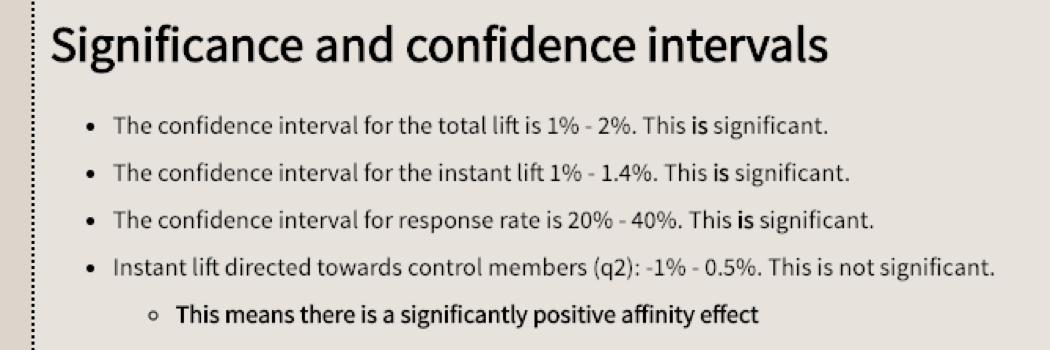}
    \caption{User Interface screenshot explaining the confidence intervals for the main result (total lift) as well as for the estimated instant lift and response rate.}
    \label{fig:significanceUI}
\end{figure*}

\begin{figure*}
    \vspace{0.5in}
    \includegraphics[trim={1in 1.5in 1in 1in}, height=7in]{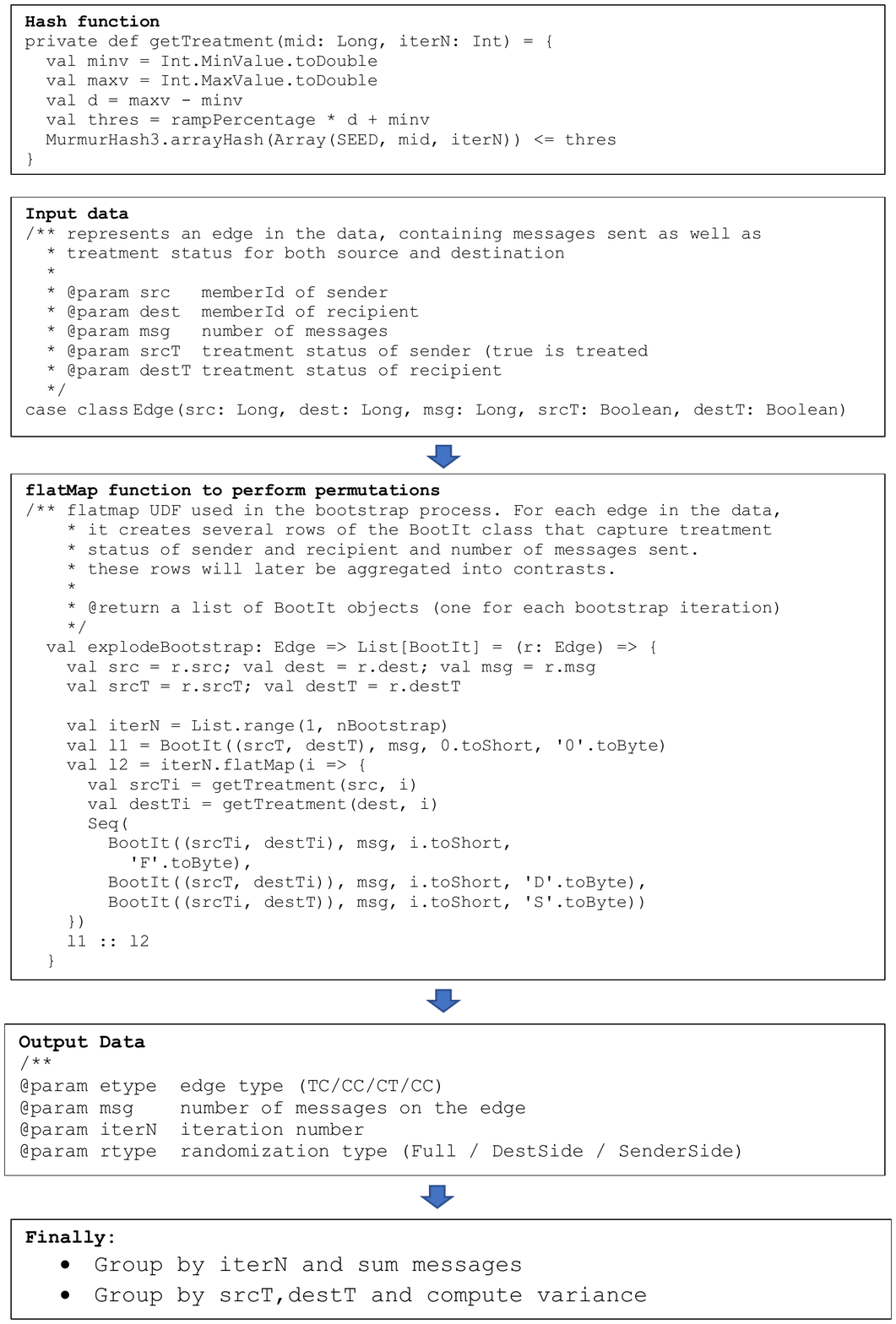}
    \caption{Algorithm summary for network-consistent permutation}
    \label{fig:algo}
\end{figure*}

\end{document}